\documentclass[fleqn,usenatbib]{mnras}

\usepackage[T1]{fontenc}

\usepackage{longtable}
\DeclareRobustCommand{\VAN}[3]{#2}
\let\VANthebibliography\thebibliography
\def\thebibliography{\DeclareRobustCommand{\VAN}[3]{##3}\VANthebibliography}

\usepackage{graphicx}	
\usepackage{amsmath}	
\usepackage{amssymb}	
\usepackage{multirow}   

\usepackage{tikz}
\usetikzlibrary{positioning,calc}
\usepackage{xcolor}
\usepackage{ulem}

\definecolor{darkcyan}{rgb}{0.0, 0.55, 0.55}

\title[Automated classification of EBs]{Automated classification of eclipsing binary systems in the VVV Survey}
\author[I.V. Daza-Perilla et al.]{
I.V. Daza-Perilla,$^{1,2,3}$\thanks{E-mail: vanessa.daza@unc.edu.ar}
L.V. Gramajo,$^{1,4}$
M. Lares,$^{1,2,4}$
T. Palma,$^{1,4}$
\newauthor
C.E. Ferreira Lopes,$^{5,6,7}$
D. Minniti,$^{8,9}$
\&
J.J. Clari\'a$^{1,4}$
\vspace*{6pt} \
\\
$^{1}$ Consejo Nacional de Investigaciones Cient\'ificas y T\'ecnicas (CONICET), Godoy Cruz 2290, Ciudad Aut\'onoma de Buenos Aires, Argentina.\\
$^{2}$ Instituto de Astronom\'ia Te\'orica y Experimental, CONICET--UNC, Argentina\\
$^{3}$ Facultad de Matem\'atica, Astronom\'ia, F\'isica y Computaci\'on, Universidad Nacional de C\'ordoba (UNC), C\'ordoba, Argentina\\
$^{4}$ Observatorio Astron\'omico de C\'ordoba, UNC, Argentina\\
$^{5}$ Instituto de Astronom\'a y Ciencias Planetarias, Universidad de Atacama, Copayapu 485, Copiap\'o, Chile\\
$^{6}$ Universidade de São Paulo, IAG, Rua do Matão 1226, Cidade Universitária, São Paulo 05508-900, Brazil\\
$^{7}$ Millennium Institute of Astrophysics, Nuncio Monseñor Sotero Sanz 100, Of. 104, Providencia, Santiago, Chile\\
$^{8}$ Departamento de Ciencias F\'isicas, Facultad de Ciencias Exactas, Universidad Andr\'es Bello, Av. Fernandez Concha 700, Las Condes, Santiago, Chile\\
$^{9}$ Vatican Observatory, V00120 Vatican City State, Italy\\
}

\date{Accepted XXX. Received YYY; in original form ZZZ}

\pubyear{2023}

\begin{document}
\label{firstpage}
\pagerange{\pageref{firstpage}--\pageref{lastpage}}
\maketitle

\begin{abstract}
With the advent of large-scale photometric surveys  of the sky, modern science witnesses the dawn of big data astronomy, where automatic handling and discovery are paramount.
In this context, classification tasks are among the key capabilities a data reduction pipeline must possess in order to compile reliable datasets, to accomplish data processing with an efficiency level impossible to achieve by means of detailed processing and human intervention.
The VISTA Variables of the Vía Láctea Survey, in the southern part of the Galactic disc, comprises multi-epoch photometric data necessary for the potential discovery of variable objects, including eclipsing binary systems ( EBs). 
In this study we use a recently published catalogue of one hundred EBs,  classified by fine-tuning theoretical models according to contact, detached or semi-detached classes belonging to the tile d040 of the VVV.
We describe the method implemented to obtain a supervised machine learning model, capable of classifying EBs   using information extracted from the light curves of variable object candidates in the phase space from tile d078.
We also discuss the efficiency of the models, the relative importance of the features and the future prospects to construct an extensive database of EBs   in the VVV survey.
\end{abstract}




\section{Introduction}

The generation of time-domain data from surveys like VISTA (Visible and Infrared Survey Telescope for Astronomy) in the Vía Láctea \citep[VVV,][]{2010NewA...15..433M, 2012AA...537A.107S, 2014Msngr.155...29H}, the VISTA Variables in the V\'ia L\'actea eXtended Survey (VVVX, \citealt{Minniti2018APlane}) and from future surveys like the Vera C. Rubin Observatory Legacy Survey of Space and Time \citep[Rubin,][]{2015HiA....16..675J} and The AURA the NASA Nancy Grace Roman Space Telescope–Rubin Synergy Working Group \citep[AURA,][]{Gezari2022} with nightly outputs ranging from \mbox{200 - 300 GB} to $\sim$~15~TB, along with other surveys such as the Catalina Real-Time Transient Survey  \citep[CRTS,][]{Drake_2009},  MACHO \citep{1992ASPC...34..193A, 1996ApJ...461...84A}, OGLE \citep{1993AcA....43...69U} and ASAS \citep{2002AcA....52..397P}, the Panoramic Survey Telescope and Rapid Response System \citep[Pan-STARRS]{2016_Chabers}, a high-cadence All-sky Survey System \citep{2018_Tonry}, Zwicky Transient Facility \citep{2019_Bellm} as well as the next generation of surveys like PLAnetary Transits and Oscillation of stars \citep{2014_Rauer} and Large Synoptic Survey Telescope \citep{2008_Ivezic} are examples of the exponential growth in astronomical data volumes \citep{2002AAS...20113406S}.
This flood of information leads to the automation of data processing and analysis.
The techniques involved in the search for patterns in large astronomical datasets are
notably dominated by supervised and unsupervised machine learning tools \citep[see, e.g., ][]{2019arXiv190407248B}.
In particular, it is usual to use information extracted from the time series data
in machine learning models for the detection, analysis, and classification of short and long period variable sources \citep{Mahabal_2019, 2020MNRAS.498.3077W}.
These tools often aim at reproducing the results obtained with human intervention
and then extend those results to large datasets through models which make the whole process fast, reliable and statistically reproducible.
Classical methods such as the "phase folding method" (PFM) and harmonic analysis (HA), among others, are implemented for the acquisition of the relevant information e.g., \citep{2007_Debosscher,  2016_Ferreira, 2017_Ferreira}. 
These methods allow the determination of the period and shape characteristics of the light curves \citep{Lomb1976,1982ApJ...263..835S, 1965ApJS...11..216L}.
Some works have combined the two mentioned methods to improve the computational efficiency and identification of variable stars \citep{2017AJ....154..231S}. 
Indeed, $\sim 75$ per cent of parameters used to characterize light curves shape are derived from the folded light curve using the period \citep{2011_Richards}. Therefore,  a combination of multiple period finding methods could potentially
increase the recovery rate \citep{2011_Dubath, 2020MNRAS.496.1730F}.
Generally, however, the works that use automatic learning models make a selection of the most informative features or else, in terms of information theory, the features chosen are the ones that have the highest entropy.
The aim is to improve the performance of the model as shown by \citet{2019MNRAS.488.4858H} using the Survey CRTS for the classification of 11 variable star types, \citet{Richards_2012} for the generation of a probabilistic classification catalogue of 28 ASAS classes and VIVACE, a catalogue of VVV variables in the bulge and disc containing hundreds of thousands of eclipsing binary systems  \citep{Molnar2021}.
These last three works are also examples of the wide spectrum of the different types of variable stars that can be classified and the challenges faced by automated methods, through which a high performance in the classification of subclasses of variable stars is hard to achieve.

The main aim of this work is to classify the subclasses of eclipsing binary systems (EBs). 
The identification of these systems is relevant since it allows to derive fundamental stellar dynamical information, as well as stellar parameters, including the masses and radii of the system components
\citep{2010Natur.468..542P, 2010AARv..18...67T}.
EBs   have also proven to be very useful in determining precise distances to nearby galaxies \citep[e.g.,][]{2006ApJ...652..313B, 2010AA...520A..74N}, improve astrophysical models \citep{2020_Carmo}, as well as in tracing the Milky Way structure \citep{2013MNRAS.432.2895H}.
In addition, the precise measurements obtained from these systems allow us to perform tests of stellar theory in the Hertzsprung-Russell diagram \citep[]{2021MNRAS.501..302A}.
In general, EBs   are classified based on their Roche lobes into three subclasses, namely, detached (D), semi-detached (SD), and contact (C) systems.
There is evidence that the performance of the EBs  classification models improves when the best features of the light curves representing each class are used and when the number of examples in the training   set is increased \citep[][]{2019MNRAS.488.4858H}.
However, this information is not conclusive enough for the classification of the SD type of EBs,  since in general the number of SD EBs  is small compared to the number of C and D systems found in the observations \citep{2006MNRAS.368.1311P}.
Therefore, our motivation is not only to design an automatic algorithm that mimics human classification but also to perform this task with a high performance, reliable, reproducible and open procedure.
For this purpose, we propose a classification model, which we refer to as a compound decision tree (CDT), built as an ensemble model of several supervised machine learning models with a voting system for the classification of the three types.
The CDT model is retrained with self-made classifications that were visually inspected. 
For the definition of the CDT model, one of the most important steps was the construction of the dataset to be used in the training   of the models.
To do so, we used an initial dataset with reliable information of the EBs,  from which we generated representative features of the subclasses and from those we selected the most correlated features with each subclass.
Our sample comprises one hundred EBs   recently published by \citet{2020PASA...37...54G}, including information on the time series, physical parameters and position of the systems.
In addition to this initial dataset, we include information about the periods of each system provided by \citet{2020MNRAS.496.1730F}.
Then, using the \textsc{feets} package \citep{2018AC....25..213C}, we extract features of the time series of each EB,  such as slope and mean. And to this set of features, we add the magnitude difference between the primary and the secondary minima.
Finally, we obtain different training  sets for the CDT model. On the one hand, we vary the set of selected features and, on the other hand, the feature selection methods.
This paper is organised as follows: In Section~\ref{sec:Data}, we detail the data used for the construction of the initial dataset. In Section~\ref{sec:method} we describe the individually implemented machine learning models used in the CDT model structure and we describe the methodology for the construction of the samples used in the training  of the models including feature generation and feature selection methods, while we analyse and discuss the results in Section~\ref{sec:results}. The compound decision tree model is described in Section~\ref{sec:CDT} and the model assessment reported in Section~\ref{S_assess}. Finally, in Section~\ref{sec:conclusions} we present the summary and main conclusions.

\section{Data}\label{sec:Data}

As in any attempt at classification, the success in discerning between classes depends mostly on the quality and quantity of the available information, so in this section we describe in detail how we constructed the dataset of EBs .
Obtaining an ensemble supervised model classifier requires, in particular, a labelled dataset including features that once combined in the model correlate with the labels.
That is, a dataset with examples of each class but with information that is important for each supervised machine learning algorithm, since each model achieves a high performance depending on the number of examples of each class and on what features are entered into it.

\subsection{Samples}

The ESO Public Survey VISTA Variables in the Via Lactea Survey$^1$ have been mapping the NIR variability ($K_s$ passband) of the Milky Way bulge and the inner southern part of the Galactic disc. The VVV survey performed observations using five NIR passbands ($Z$, $Y$, $J$, $H$, and $K_s$) plus the variability campaign on the $K_s$ passband over the period 2010–2017 (on average about 100 $K_s$ epochs per field). CASU (Cambridge Astronomy Survey Unit) with the VIRCAM pipeline v1.3 \citep{2004SPIE.5493..411I, 2004SPIE.5493..401E}$^2$ is used to produce the reduction and aperture photometry. The readers can find a detailed description of the VVV survey in many papers published in the last few years \citep[e.g.,][]{2012AA...537A.107S, 2014Msngr.155...29H, 2014IAUS....301..395C}. In particular, this paper is the second of a series of studies about the EBs included in the VVV survey.

In the first paper of this series \citep{Gramajo2020}, the VVV tile d040 (centred at RA2000 $= 11^{h}58^{m}14.16^{s}$, DEC2000 $= -62^{\circ} 48'15.12''$, corresponding to the Galactic coordinates: l $= 296.8962^{\circ}$, b $= -0.5576^{\circ}$) was explored and approximately \mbox{$1.6 \times 10^6$} sources were found, from which \citet{Gramajo2020} pre$-$selected 3100 variable sources by means of the Stetson variability statistics \citep{1996PASP..108..851S}.   
Then, they visually inspected these 3100 variable sources after the acquisition of the phase-folded light curves with their preliminary periods through the Generalised Lomb$-$Scargle and the Phase Dispersion Minimisation algorithms, respectively \citep{2009AA...496..577Z, 1978ApJ...224..953S}, and obtained a sample of 400 sources that were not spurious signals. Finally, they interactively redefined the periods to optimise the light curve fits by using a non-linear Fourier fit and obtaining average apparent magnitudes and total amplitudes in the $K_s$ band. From a visual inspection of the phased light curves, a final sample of one hundred EBs   was gathered, where 50 of them are detached, 13 are semi-detached and 37 are contact binaries.

\citet{2020PASA...37...54G} determined the physical and geometrical parameters of the EB  sample by means of the \textsc{PHOEBE $1.0$} code \citep{2005ApJ...628..426P}, as a result of which the studied sample provides information on the effective temperatures of the two components ($T_1$ and $T_2$), the mass ratio ($M_2/M_1$), the orbital inclination ($i$) of the system, the orbital eccentricity value ($e$), the relative size of the two components ($R_1/R_2$) and the accuracy of the fit by the $\chi^2$ value, which measures the discrepancy between the observational data and the adopted model used to determine the physical parameters. 

In the cited catalogue, we included the periods provided by \citet{2020MNRAS.496.1730F} and selected those with values in the interval $0.4-15.2$ days. This condition reduced the total number of systems to $96$, which we now refer to as "initial D", whose classes are intrinsically unbalanced, with 49 Detached, 35 Contact and 12 Semi-Detached systems.

\begin{figure*}
	\includegraphics[width=1\textwidth]{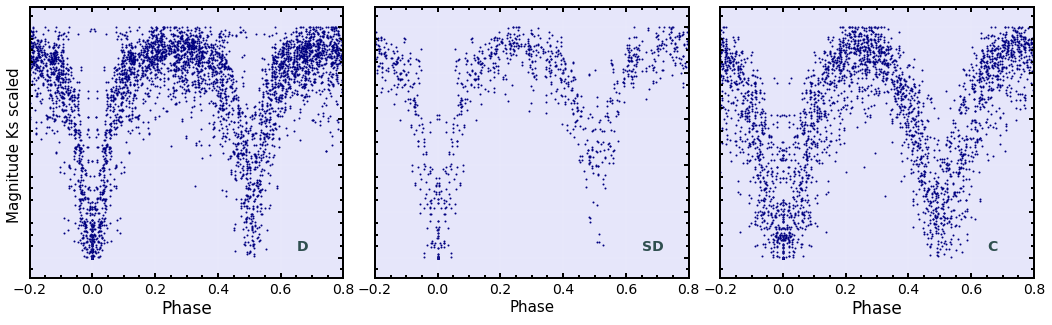}
    \caption{Composite EB light curves, separated by class, which are scaled from 0 to 1 and with the primary minimum located at zero. These composite light curves are made using the individual measurements of 49 Detached systems (left panel), 12 Semi-Detached systems (middle panel), and 35 Contact systems (right panel). }
	\label{fig:fit_D_SD_C_CL}
\end{figure*}

In Figure~\ref{fig:fit_D_SD_C_CL}, we show the light curves measurements for each system, distinguished by class (systems D, SD and C in the left, middle and right panels, respectively). 

The magnitudes of each light curve were scaled to [0, 1], the primary minima of each light curve being located at the phase equal to zero.

Finally, we use a second sample that is also found in the outermost region of the Galactic disc and in tile d078 centred on RA2000 $= 12^{h}00^{m}07.584^{s}$, DEC2000 $= -61^{\circ} 44'3.84''$, corresponding to the Galactic coordinates: l $= 296.89636^{\circ}$, b $= 0.53458^{\circ}$, where the initial sample contains 41,508 candidates. For this sample, we used the same condition for the periods as for tile d040.


\section{Methods}\label{sec:method}

Among the popular modern machine learning (ML) algorithms, different types are usually considered: supervised, unsupervised, semi-supervised and reinforcement learning models.
For supervised models, the training   data is used expecting an output, i.e., it is known in advance how it can be labelled or classified, while in unsupervised learning models the training   data is used without an expected output, i.e. no label is available.
Semi-supervised learning models use both labelled and unlabelled train data, usually a small amount of labelled data together with a large amount of unlabelled data.
Reinforcement learning models are used to determine what actions a software agent should choose in a given environment in order to maximise some notion of ``reward'' or accrued reward.
We used supervised models in this work, since we wanted to solve a classification problem and we have labelled training  sets. 
Specifically, we used five of the best known supervised learning models, both individually and in a combination of some of them to create a single model which we call the ``compound decision tree model''.
There is a large amount of material explaining ML methods in detail. However, find below a description of each of the individual models:

{\it Decision tree (DT)}: they are algorithms for classification using successive partitions. They are appropriate for large datasets; one their advantages  is their descriptive character. This characteristic which allows us to easily understand and interpret the results obtained by the model.  They select a variable from the dataset and then a specific value for this variable. This gives us a better classification of the data: the algorithms reveal complex shapes in the data structure that cannot be detected with conventional regression methods \citep{2017_Breiman}.

{\it Random forest (RF)}: random forests are a combination of tree predictors so that each tree depends on the values of a random vector independently sampled and with the same distribution. The parameters of a random forest are the variables and thresholds used to split each node learned during training  .

As regards the generalisation error of the forests, it converges to a limit as the number of trees in the forest becomes large; this will be good when the model has learned the pattern between the input data and the results. Therefore, the model can be trusted to perform well with new data; at least as long as the new data come from a similar distribution. The generalisation error of a forest of tree classifiers depends on the strength of the individual trees in the forest and on the correlation between them \citep{2001_Breiman}.

{\it K-nearest neighbour (KNN)}: the neighbours-based classification is a type of instance-based learning or non-generalizing learning: it does not attempt to construct a general internal model but simply stores instances of the training   data. Classification is computed from a simple majority vote of the nearest neighbours of each point: a query point is assigned the data class which has the most representatives within the nearest neighbours of the point \citep{Vaidya1989AnOnProblem}.

{\it Linear support vector classification (LSVC)}: given a set of training  examples with each class marked as belonging to one of two categories, the support vector model performs a non-probabilistic linear binary classification.
The classification by support vector assigns  the examples in the training set  to points in the space to maximise the width of the gap between the two categories, by defining a hyperplane that separates them. The new examples are then assigned a category according to which side of the separating hyperplane they are on \citep[][]{LinearSupportVector}.

{\it Neural networks (NN)}: a standard neural network consists of many simple, connected processors, called neurons, each of which produces a sequence of real-valued activations. A NN contains an input layer, one or more hidden layers and an output layer. Each artificial neuron is connected to other neurons and has an associated weight and threshold.

Learning or weight assignment consists of finding the weights that cause the NN to exhibit the desired behaviour, such as driving a car. Depending on the problem and how the neurons are connected, such behaviour may require long causal chains of computational stages, with each stage often non-linearly transforming the output of the neurons \citep{SchmidhuberJ_2015}.

\subsection{Generation of features}
\label{sec:features} 
The light curves of binary systems include information such as period, amplitude and shape that depend on the physical parameters and characteristics of the system.
Then it should be possible, in principle, to distinguish the different EB  types on the basis of the geometrical characteristics of the curves.
Since this is not a physically motivated model, the properties of the system and their influence on the classification are introduced through the features of the training  dataset.
Therefore, extracting the information from these time series and selecting the most relevant ones generates a much more accurate model for classification. 

To extract both statistical and shape features from the light curves (mean, scatter, slope, etc.), we used the \textsc{feets} code   \citep{2018AC....25..213C}. \textsc{feets} takes as input the magnitude, in our case the $K_s$ magnitude, its error and the time (Julian Date). In addition, we calculated the difference in height $\Delta m$ between the primary minimum and the secondary one, and included the different periods provided by \citet{2020MNRAS.496.1730F}. As a result, we obtained 64 features of the time series of each EB  which, in addition to the periods, make a total of 69 features for the 96 sources.
In Table~\ref{tab:table_1}, we show the 69 features differentiated in 5 types: the first one is related to the form or shape of the curve (F) and includes all the values obtained from the distribution of magnitudes.
The "Frecuency type" (Fr) set of features contains values resulting from the modelling of the time series in frequency space.
The time--related features (T) comprise all the periods that were provided and obtained through \textsc{feets}.
The "statistics type" set (St) contains the statistical values of the flux distribution determined through the equation \ref{equ:flux}:

\begin{equation}
 \hspace{3 cm} F = 10^{-0.4 * mag}
\label{equ:flux}
\end{equation}

\noindent and the last set of the table includes three values calculated from the structure function with variable equal to the magnitudes $K_s$ as a function of time.  It should be noted that it is in the training  process that a selection of the most relevant features is made.

\begin{table*}
	\centering
	\caption{Initial set of form, statistical, frequency, temporal and physical features of the light curves of the 96 EBs  . Most of these features were obtained with \textsc{feets}. The rest corresponds to the different periods and to the difference in amplitude between the minima of the light curve ($\Delta m$). In bold the most important features according to the MI  method.}
	\label{tab:table_1}
	\begin{tabular}{lllll} 
		\hline
	    \multicolumn{4}{c}{Features}\\
	    \hline
		\hline
		Form  & F1: MedianBRP & F2: Skew & F3: Beyond1Std  & F4: Gskew\\
		(magnitude)   & F5: StetsonK  & F6: AndersonDarling  
               & F7: MedianAbsDev  & F8: SmallKurtosis\\
               & F9: StetsonK AC & F10: PercentAmplitude & F11: Psi CS  & F12: Psi eta\\
               & F13: Meanvariance  & F14: Amplitude  & F15: CAR mean  & F16: CARs sigma\\    
               & F17: CAR tau   & F8: Autocor length & F19: Con
               & F20: Eta e\\
               & F21: LinearTrend   & F22: MaxSlope  & F23: Mean 
               & F24: PairSlopeTrend\\ 
               & F25: Q31 & F26: Rcs & F27: SlottedA length & F28: Std\\
               & F29: $\Delta m$\\
	\hline
	Frequency & Fr1: Fr1 harmonics A0 & Fr2: Fr1 harmonics A1 & Fr3: Fr1 harmonics A2 & Fr4: Fr1 harmonics A3\\
    & Fr5: Fr1 harmonics $\phi$ 0 & Fr6: Fr1 harmonics $\phi$ 1 & Fr7: Fr1 harmonics $\phi$ 2 & Fr8: Fr1 harmonics $\phi$ 3\\
    & Fr9: Fr2 harmonics A0 & Fr10: Fr2 harmonics A1 & Fr11: Fr2 harmonics A2 & Fr12: Fr2 harmonics A3\\
    & Fr13: Fr2 harmonics $\phi$ 0 & Fr14: Fr2 harmonics $\phi$ 1 & Fr15: Fr2 harmonics $\phi$ 2 & Fr16: Fr2 harmonics $\phi$ 3\\
    & Fr17: Fr3 harmonics A0 & Fr18: Fr3 harmonics A1 & Fr19: Fr3 harmonics A2 & Fr20: Fr3 harmonics A3\\
    & Fr21: Fr3 harmonics $\phi$ 0 & Fr22: Fr3 harmonics $\phi$ 1 & Fr23: Fr3 harmonics $\phi$ 2 & Fr24: Fr3 harmonics $\phi$ 3\\
    \hline
    Temporary & T1: Period LSG & T2: Period PDM & T3: Period LS & T4: Period STR\\
          & T5: Period KFI & T6: Period fit & T7: Period PAN\\
    \hline
    Statistics  & S1: FluxPercentile35 & S2: FluxPercentile65 & S3: FluxPercentile50 & S4: FluxPercentile20\\ (flux) & S5: FluxPercentile80 & S6: PercentDifferenceFluxPercentile \\
    \hline
    Structure function & SF1: Structure index 31 & SF2: Structure index 32 & SF3: Structure index 21\\
    \hline
	\end{tabular}
\end{table*}

\subsection{Split and pre-processing}
\label{sec:pre-processing}

As a first step, we divided the labelled EB  sample into two sets, comprising 80 per cent and 20 per cent of the total sample for train and test, respectively.
In the training   of the Random Forest and Neuronal Networks models, the set with 80 per cent of the total sample was taken and divided into two, one for the training   of the parameters (training   set) and the other for the choice of the hyperparameters (validation set). The sizes of the train and test sets for the initial D sample are shown in the first row of Table~\ref{tab:table_2}.

\begin{table}
\centering
\caption{In the development of the classifiers, initial D is divided into two sets (train and test) for model training  , hyperparameter selection and model testing, respectively.  The tuples in each field of the table indicate the number of EBs   in each set in the first component of the tuple. In the second component of the tuple, the number of features describing each EB  is indicated. The first row shows the dimensions of the initial data, while the second row displays the dimensions of the curated and pre-processed data.}

\begin{tabular}{lccc}
\hline\hline\noalign{\smallskip}
\!\! dataset & \!\!\!\!Training [\#EBs  , \#features] & \!\!\!\! Test [\#EBs  , \#features] \!\!\!\!\\
\hline\noalign{\smallskip}
\!\!Initial data &  (D:38 | C:28 | SD:10, 69)      & (D:11 | C:7 | SD:2, 69)\\

\!\!Processed data & (D:38 | C:28 | SD:10, 35)     & (D:11 | C:7 | SD:2, 35)\\
\hline\noalign{\smallskip}
\end{tabular}
\label{tab:table_2}
\end{table}     

We used the min-max scaling method for each of the features. The reason for this pre-processing is that many of the ML models use the distance between the points in the feature space and, if they are not scaled to the model, could perform poorly given the large variability of the feature distribution. 

Out of the 69 features obtained, those with non-zero variance were taken, since constant values are not important in the training  of ML models. Therefore, the feature set was reduced to 64, discarding three components of the Fourier decomposition: Fr1 harmonics  $\phi_0$, Fr2 harmonics  $\phi_0$, Fr3 harmonics  $\phi_0$ and two parameters which provide a natural and consistent way of estimating a time scale through a continuous time auto regressive model applied to time series, CAR tau and CAR sigma \citep{2002brockwell}.

\subsection{Selection of Features}
\label{sec:Selection of Features}

Taking into account that we aimed at reaching a predictive classification modelling problem with numerical input and categorical output variables and that the performance of supervised ML models depends on the number and selection of the features, we conducted an univariate analysis study, which consists of estimating the importance of each attribute individually along with the class of the attribute.
In particular, we used the correlation techniques between the input variables and the target recommended by \citet{Jason_2020} for this type of classification problems: ANOVA, mutual information (MI ) and $\chi^2$.
With each technique, we will generate a list of scores for each feature, which depends on the correlation that each technique finds between the feature and the class.
Therefore,  the features with low entropy for the classification of EBs   will occupy the lowest places in the ranked list of features of each method.
Consequently, to determine in the initial D the number of features and to select the subset of features, we observed the performance obtained by the ML models, when using all features and when selecting feature sets consisting of the first 10, 35 and 50 features resulting from the ANOVA, MI  and $\chi^2$ techniques, respectively.
Thus, we could study the impact on model performance when using a small sample, half, or almost all the features of the initial set. 

\subsection{Balance of classes}

In general, training   ML models with imbalanced samples in classes affects classification, since having many examples of a particular class means that the models learn to classify that class very well and not the minority ones \citep{2012_Mao}.
Therefore, we included balanced sets in our study.
The choice of how to balance the data depends on the classification problem, on the number of examples of each class and on the model. Thus, in order to build a balanced sample, we implemented two methodologies on the training   set.

The first method is \textsc{SMOTETomek}, which consists of combining oversampling in the minority class (in our case, the C and SD classes) and undersampling the majority class (D class).
This method uses two separate techniques: \textsc{SMOTE} (Synthetic Minority Over-sampling Technique, \citet{Chawla_2002}) and \textsc{Tomek links} \citet{Ivan_1976}.
\textsc{SMOTE} performs the oversampling by creating "synthetic" examples in the "feature space". Minority classes are oversampled by taking each sample of the minority class and introducing synthetic examples along the line segments joining any/all k nearest neighbours of the minority class.
The other technique is \textsc{Tomek}, originally proposed by \citet{Ivan_1976}.
It consists of a one-on-one scheme, which generates samples from a given dataset that preserves or improves the performance that would be obtained with the initial D. Therefore, if we implement this method in case of having two imbalanced classes, the number of examples of the majority class decreases a little and the number of examples of the minority class increases until it reaches the same number as the class that previously had the highest number of examples.
The second method is \textsc{RandomOverSampler}, a.k.a. Random Over-Sampling Examples : \citep[ROSE,][]{Menardi_2014}.
This procedure only oversamples the minority class by randomly choosing samples through smoothed bootstrap.
The reason for generating samples using different methods to remove the imbalance lies in the fact that the distribution of features in each sample will be different, which affects the performance of the models.

\subsection{Hyperparameters}

The hyperparameters of a model are the values of the configurations used during the training   process and depend on each particular model.
Those values are not generally obtained from the data and therefore usually provided by the data scientist.
Hyperparameter tuning relies more on experimental results than on theory, so the best method to determine the optimal settings is to try many different combinations and evaluate the performance of each model.
However, evaluating each model only on the training   set can lead to overfitting, which is one of the most fundamental problems in ML. Therefore, to avoid overfitting, we used cross-validation, in particular the K-fold cross-validation method that divides our training   set into a number of K subsets, called folds.
We then iteratively fit the model K times, each time training   the data on K-1 of the folds and evaluating on the K-th fold (called validation data). At the very end of training  , we averaged the performance on each of the folds to come up with final validation metrics for the model. For hyperparameter tuning, we performed many iterations of the entire K-fold cross validation process, each time using different model settings. We then compared all the models, selected the best one, trained it on the full training   set and then evaluated it on the test set.

\subsection{Metrics}

Since in our study we used balanced and imbalanced datasets that present an inherent imbalance in classes, we preserved in the procedure that imbalance in the development and evaluation. For that reason, in our selection of metrics, we included as a decisive metric the micro average of the F1-value ($F1_{mic}$), which does take into account the mentioned class imbalance, in addition to the precision (P), recall (R), F1-score (F1) and support (S) metrics and also the confusion matrices.

\subsection{Workflow}

\begin{figure*}
    \centering
    \includegraphics[width=2\columnwidth]{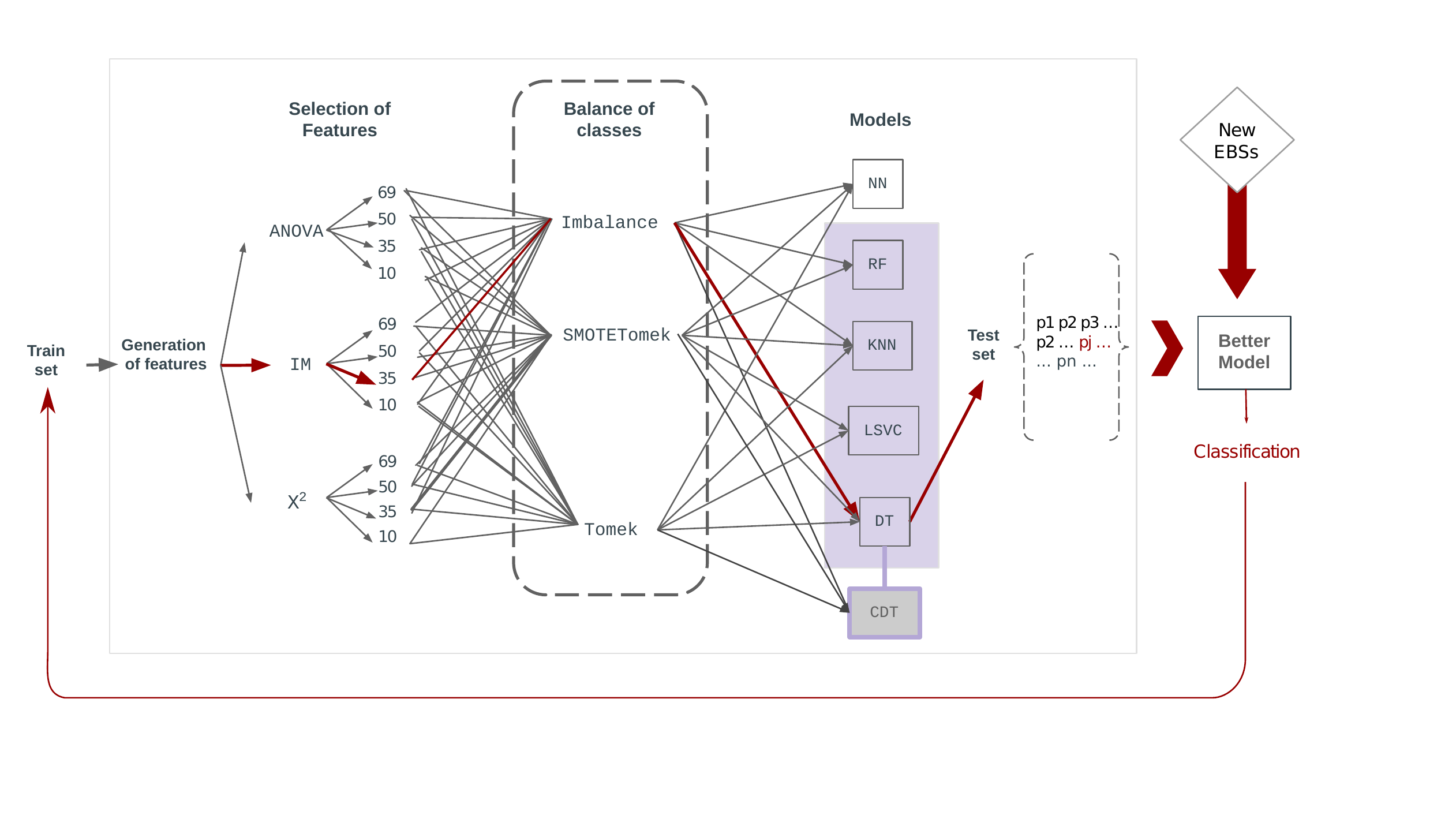}
    \caption{Schematic of the process to determine the best model for EBs   classification. classification. The generation of different training sets is shown inside the box.  A training   set is constructed through the choice of one of the feature selection methods: ANOVA, MI  and $\chi^2$, the number of features and one of the balancing methods: SMOTETomek and Tomek. The red line indicates one of these combinations. The evaluation and improvement of the model using a new dataset is schematised outside the table. }
    \label{fig:esquema}
\end{figure*}

Taking into account the methodology described in this section, we generated a workflow to determine the best model. Firstly, we took the training   set and generated the features for each EB. Then, we used a feature selection method and took some or all of the most important features. 
Next, we used the imbalanced or balanced set. After that, we trained the models with the constructed set and compared the performance on the validation set. This process was carried out with all combinations of features and balanced or imbalanced sets. Finally, we compared the performances and chose the best model. A schematic of the process is presented in Figure~\ref{fig:esquema}. 

\section{Results}\label{sec:results}

\begin{figure*}
	\includegraphics[width=2\columnwidth]{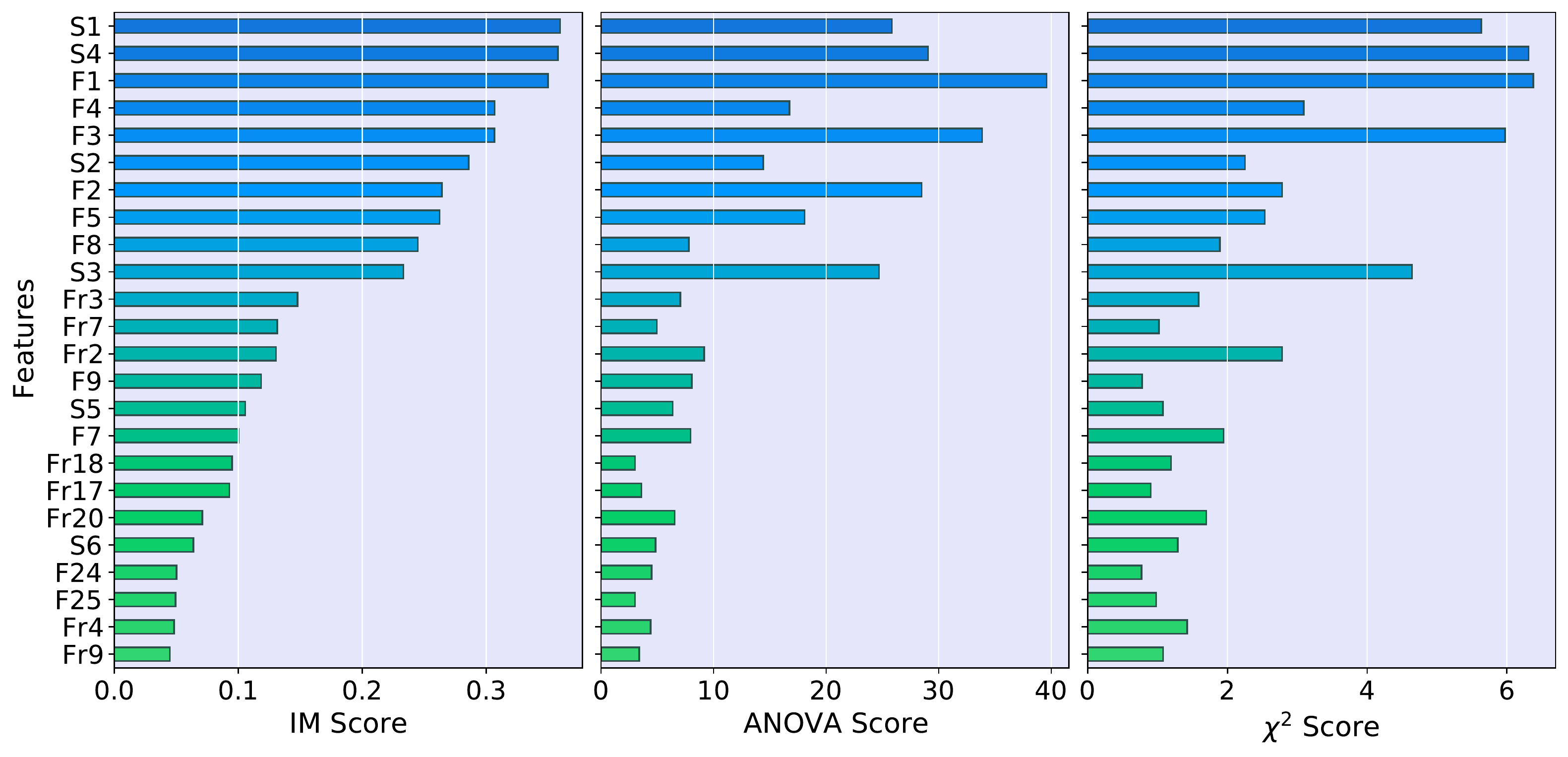}
    \caption{Bar chart of the scores of the 24 most correlated features with the classes calculated from the features selection, Mutual Information (MI), ANOVA and $\chi^2$ methods.  Results obtained by applying the methods on the train sample with D, C and SD classes.}
	\label{fig:figure_2}
\end{figure*}

\begin{figure}
	\includegraphics[width=\columnwidth]{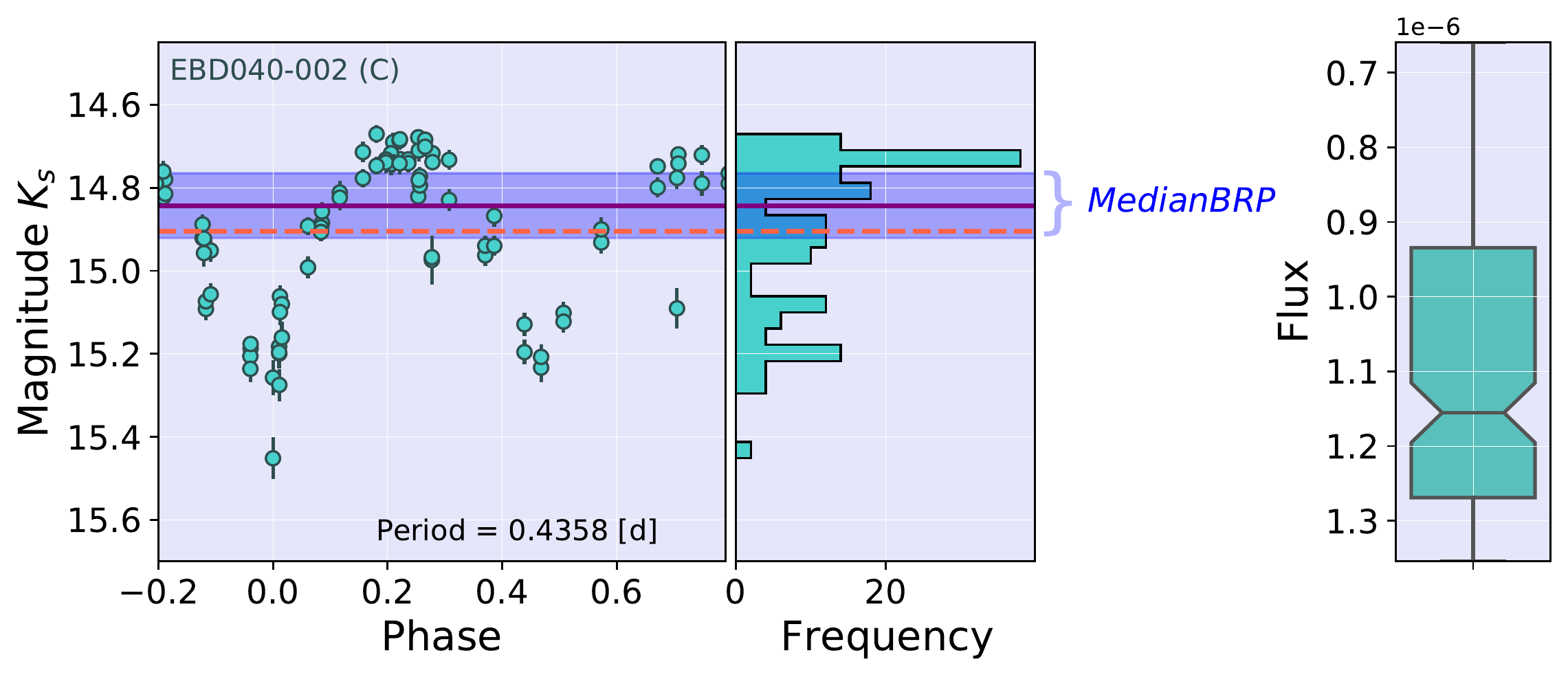}
	\includegraphics[width=\columnwidth]{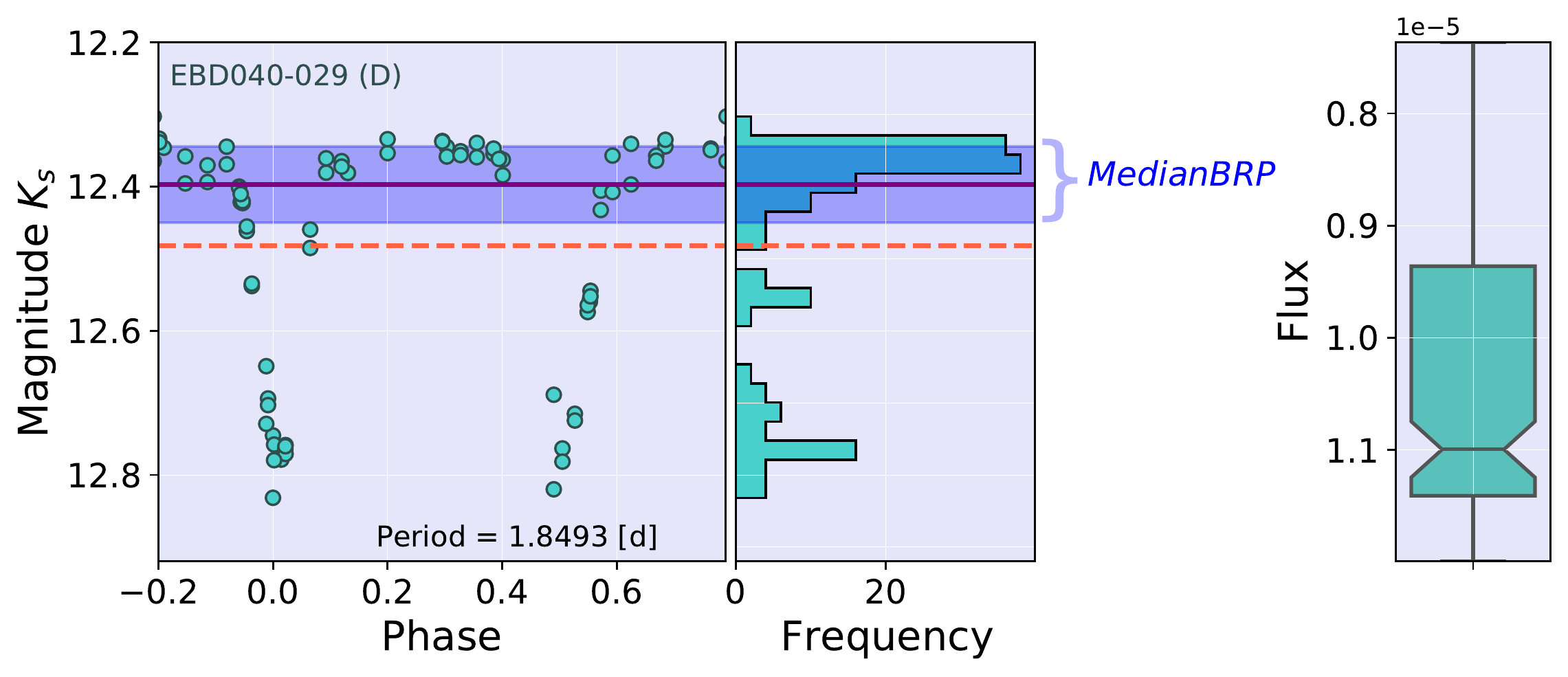}
    \caption{Light curves together with the $K_s$ magnitude distribution of two EBs.  Contact in the upper panel and detached type in the lower one. The violet band shows the values used to determine MedianBRP. The solid and dashed lines indicate the median and mean, respectively.}
	\label{fig:CL_Features}
\end{figure}

Based on the methodology described in Section~\ref{sec:method}, we generated the features of the light curves and took the cured train dataset, i.e., the dataset with 64 features scaled between [0, 1] without the 5 features with variance equal to zero. The resulting dimensions of the training  and test sets are shown in the second row of Table~\ref{tab:table_2}. With the training  set and the DT, RF, KNN and LSVC models, we searched for the features that give the best performance.
To this aim, we ran several experiments combining the different models with different inputs and used the curated dataset of 64 features, the dataset with some periods, and datasets of 10, 35 and 50 features with the highest scores according to ANOVA, MI  and $\chi^2$ techniques.
Although the performance of the ML models depends on the features that enter the model (in this case the features that are extracted from the light curves), here the performance of the models with different features varied in most cases by hundredths and in very few cases by tenths, mainly because the feature sets generated by the three feature selection techniques have similar results in the score of the most important features for the classification of EBs .
For the three feature selection techniques, 24 out of 35 and 8 out of 10 features with the highest performance match one another in the three methods.
Figure~\ref{fig:figure_2} compares the scores of the 24 features that the three methods have in common. 
In particular, the first features are "form" and "statistical" measures extracted from the light curve.
One of the features we found most relevant in the 3 methods is MedianBRP, as shown in Figure~\ref{fig:CL_Features} for two light curves of C and D EBs .
This one is related to the shape of the curve and indicates the fraction ($<= 1$) of photometric points within the amplitude/10 of the median magnitude.
The light curve of contact binary at maxima have usually more points if compared to the detached EBs   curves because there is a higher probability of finding more measurements of this type of binary system when their system components are overlapping, since the times of this event are shorter compared those of D EBs .
On average for our training   set, we conclude that the median value of this measurement for types C, D and SD EBs  is 0.61, 0.28 and 0.52, respectively. Another form feature we found important was Beyond1Std, which constitutes the percentage of points beyond one standard deviation from the weighted mean. It gives a measure of the number of observations at the extremes of the photometric point distribution. This fact is good for distinguishing the D EBs   from the other classes, since at the extremes of the magnitude distribution the D EBs   show a larger fraction of points.
This is due to the fact that they have the contribution of two minima, while the SD and C EBs   receive, in general terms, the contribution of a single minimum.
Regarding the statistical features, we noticed that some percentile of fluxes are significant for the distribution between the classes, such as {\it Flux Percentile 20} and {\it Flux Percentile 35}, values calculated from the percentiles $F_{5,95}$, which is the difference between 95th and 5th percentiles, so {\it Flux Percentile 20} is $F_{40,60}/F_{5,95}$ and {\it  Flux Percentile 35} is $F_{32. 5,67.5}/F_{5,95}$.
These values in the training   set distinguish well between EBs   classes C and D, since the values for the SD class of EBs   are more similar to class C, as shown in Table~\ref{tab:table_3}, where the mean of each of these features for the three EBs   classes is displayed.
However, we found that not all features of C and SD are similar to each other, in particular the LS period (Lomb Scargle period) of SD EBs   whose values are more similar to the D EBs   class (see Table~\ref{tab:table_3}). 
Although it is not the most correlated feature with the classes for any of the three methods, it is important to keep it in the input of the models as combining it with the more informative features will help to distinguish between the C and DS of EBs   classes.

\begin{table}
\centering
\caption{Mean values of the five important features for the three feature selection methods in the training   set.}
\begin{tabular}{llll}
\hline\hline\noalign{\smallskip}
\!\! Features & \!\!\!\! D & \!\!\!\! C & \!\!\!\! SD \!\!\!\!\\
\hline\noalign{\smallskip}
\!\! Median BRP & 0.29 $\pm$ 0.10 & 0.58 $\pm$ 0.15 & 0.50 $\pm$ 0.13\\
\!\! Beyond 1Std & 0.31 $\pm$ 0.07 & 0.20 $\pm$ 0.05 & 0.20 $\pm$ 0.05\\
\!\! Flux Percentile   20 & 0.22 $\pm$ 0.07 & 0.09 $\pm$ 0.07 & 0.11 $\pm$ 0.05\\
\!\! Flux Percentile  35 & 0.39 $\pm$ 0.09 & 0.20 $\pm$ 0.13 & 0.21 $\pm$ 0.06\\
\!\! Period LS & 1.08 $\pm$ 1.76 & 2.06 $\pm$ 2.62 & 0.89 $\pm$ 0.04\\
\hline\noalign{\smallskip}
\end{tabular}
\label{tab:table_3}
\end{table}

Our analysis shows that the set with the highest importance is practically the same for the three feature selection methods, so we can fix one of these methods and take a smaller set of features that is reliable at learning any supervised learning model. Therefore, we chose the MI  method to select the 35 most informative features for classification to build the classifier model, marked in bold in Table \ref{tab:table_1}. 
In addition, the analysis also showed that the set of 35 features chosen is good for classifying between classes C and D but not good enough for classifying SD EBs . This problem generally arises in the classification of EBs   given the small number of examples we have of this type of EBs   and the similar features that SD has with C and D EBs  . We understand then that the model must be prepared to receive an imbalanced input with SD as the minority class. 
It is therefore important to take into account in the training   that the model needs to input a good number of examples of each class so that it does not learn to predict only the class with more examples.
We also analysed how the performance of the models behaves when inputting the imbalanced and balanced training   sets with the \textsc{SMOTETomek} and \textsc{RandomOverSample} methods. As a result, the performance using these three training   sets does not vary significantly so we decided to use all three sets in the classifier training . 

\begin{figure}
	\includegraphics[width=0.49\columnwidth]{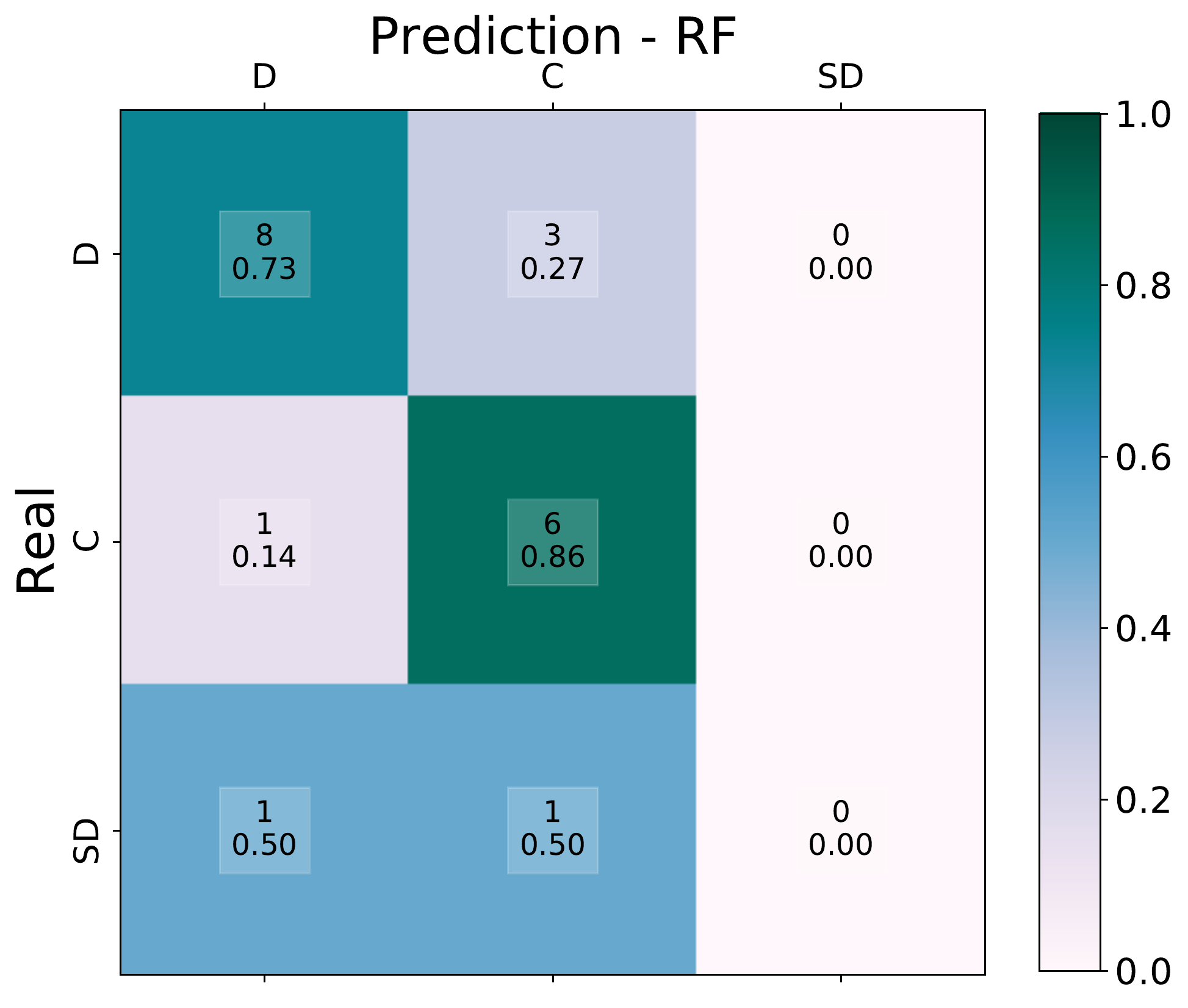}
	\includegraphics[width=0.49\columnwidth]{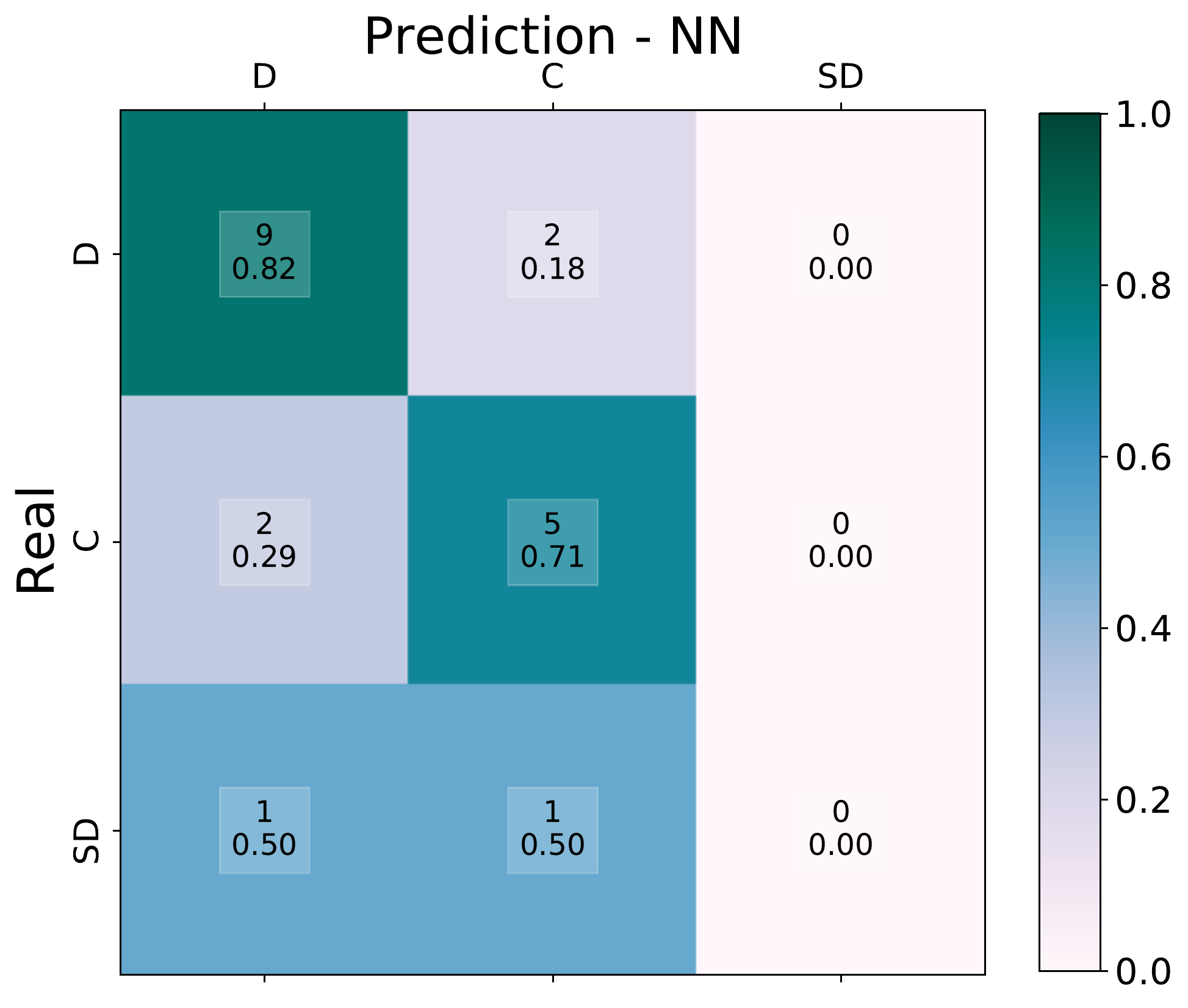}
    \caption{Confusion matrix visualisation of the performance. The cells indicate the fraction of elements in rows that are classified with the labels in the columns, so that the elements in the diagonal are correctly classified. {\it Left panel}: performance of the RF model. {\it Right panel}: performance of the NN model.}
	\label{fig:NN_RF}
\end{figure}

As regards the choice of the best classifier model, we found that the performance variations among the base models is small, regardless of introducing different sets of features, with different class balance strategies and with a variation in some hyperparameters.
For example, we considered as variables the number of layers of the NN or the depth of the trees in the RF model.
Then the choice of the model is not straightforward, even more so when none of the models achieves a high performance in the classification of the SDs. Figure~\ref{fig:NN_RF} shows the performance of two of the models that tend to have high performance given that they have more parameters than the other models, RF and NN, in the test set with 2 SDs.

\section{Compound decision tree (CDT)}\label{sec:CDT}

The performance in classifying the SD EBs   of each model with different inputs has not been high compared to the classification of D and C EBs.  However, we observed that in some experiments the models classified the SDs asynchronously, i.e., in some cases an SD EB  is classified by one or several models but not by all the others.
Since there are cases where the features of the SD EBs   can be very similar to the C EBs   and there are other cases where they can be very similar to the D EBs , we decided to make a model composed of different models with a voting system, which we called Compound Decision Tree (CDT). 
This model is mainly composed of three models ones, $M_1$, $M_2$ and $M_3$, following an ordering of execution and meeting some requirements. The CDT is structured as follows: when a data $x$ is entered, the first model to be used model is M1, which classifies between class D and C of EBs .

The next step depends on the classification resulting from M1: if the classification is $x = D$, then $x$ enters model M2, which classifies between class D and SD. Otherwise, if $x = C$, then $x$ enters model M3 and $x$ is classified into C or SD.
Each $M_i$ model uses 4 algorithms, KNN, DT, RF and SVML, which were trained on 3 EBs  samples with D, SD and C classes.
The first sample, $s_1$, has the natural imbalance of EBs  , i.e., a higher number of D class EBs  , followed by C class EBs   and a very small number of SD class EBs . Sample $s_2$ is sample $s_1$ with increased data from the minority class, i.e. class SD and sample $s_3$ is sample $s_1$ with a decrease of examples from classes D and an increase of class C and SD.
Therefore, each $M_i$ model is formed by 4 algorithms trained with the different samples $s_1$, $s_2$ and $s_3$, which leaves a total of 12 sub-models since the parameters of each algorithm change when different samples are used in the training  .

\begin{figure}
    \centering
	\includegraphics[width=0.7\columnwidth]{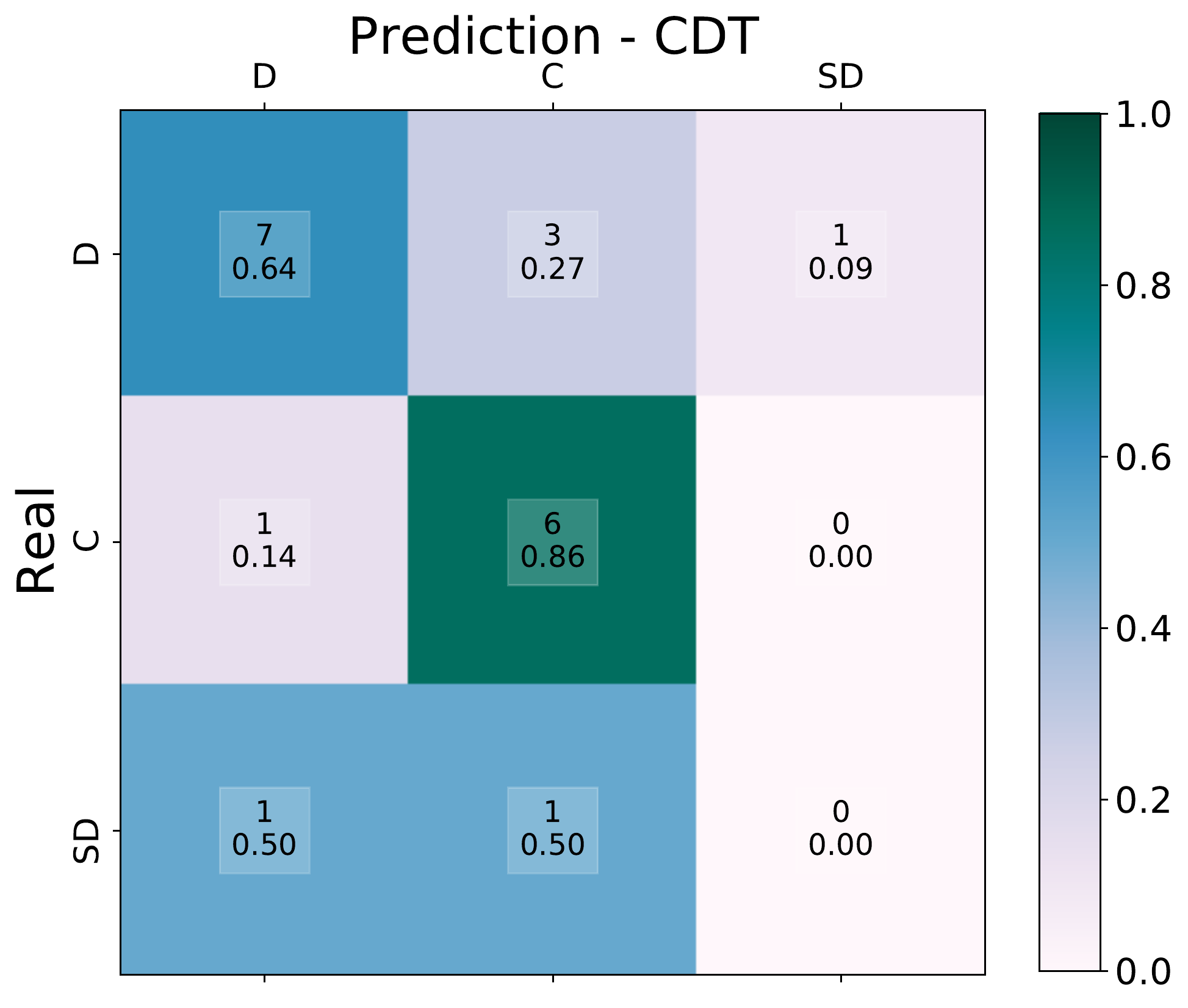}
    \caption{Confusion matrix visualisation of the performance in the model CDT. The cells indicate the fraction of elements in rows that are classified with the labels in the columns, so that the elements in the diagonal are correctly classified.}
	\label{fig:CTD76}
\end{figure}

These $M_i$ models consisting of 12 sub-models perform classifications between two classes, e.g., $M_1$ classifies between C and D, it will be C if the majority of the sub-models had this result, otherwise the classification will be D and the majority of the sub-models will have classified the input as C. If the classification or vote of the sub-models gives a tie in the M1 model, the EB  will be discarded.
In the case of a tie vote in the M2 or M3 models, the EB  class will be D or C, respectively. 
The model was tested on the same test set as the individual models and is not outperforming (see Figure~\ref{fig:CTD76}).

\section{Assessment and improvement of models for automatic classification}\label{S_assess}

\begin{figure}
	\includegraphics[width=1\columnwidth]{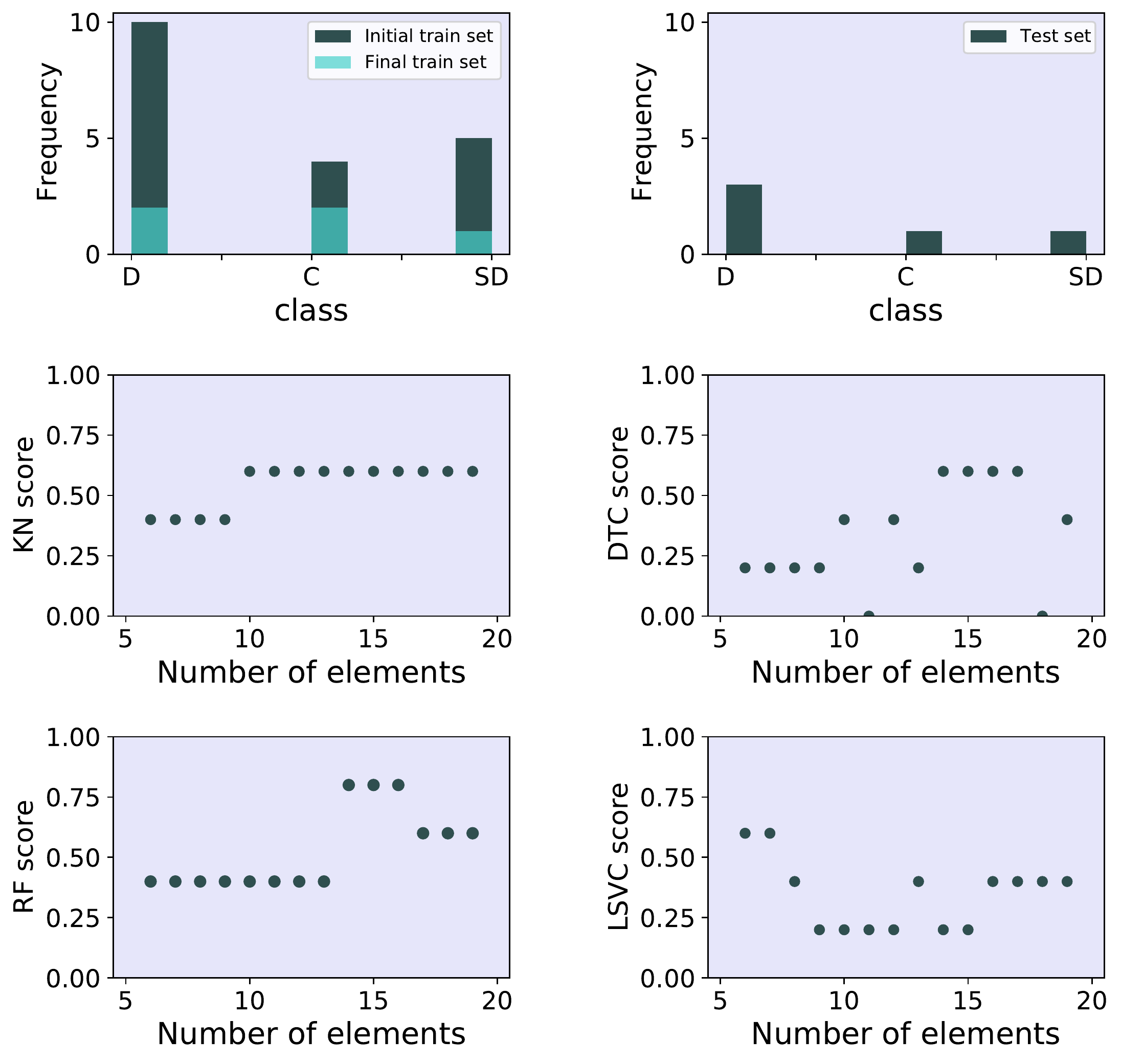}
    \caption{The first row shows the balance of the EB  classes in the training  and test sets. The second and third rows correspond to the performance achieved by the models when the number of members in the training   sample is varied.}
	\label{fig:df3_jackknife}
\end{figure}

The results strongly depend on the quality of the dataset, the number of examples of each class of EBs  and the size of the set. To study the impact on performance according to these variables, we took the 96 EBs  and generated 4 subsamples of 24 elements , 20 EBs  are used as training   set and 4 as test set. With each sample, we observed how the results vary when the training   samples have one element removed to reach a set of 6 elements. Figure~\ref{fig:df3_jackknife} shows the results with one of the four samples: this particular sample has in the test set all three classes of EBs  . 

Here it can be observed that despite having a small number of elements in the training  sample, not all models perform poorly. Likewise, we find that although some models have almost the entire training   set, they can perform at zero.  This is the case when the training set  has very few examples of a class and many in the test set. Therefore and in order to have higher reliability in classifying EBs automatically on any VVV tile, we start from a time series set of 41508 variable objects on tile d078 to classify them into D, SD and C EBs through the RF, NN and CDT models. 
The information of the new candidates EBs  to be classified must follow the same process as the dataset used to train the model. So from the time series of each EB we obtained by means of \textsc{feets} the 35 features with high entropy according to the MI  method. This step reduced the number of objects to \textsc{38538}, as \textsc{feets} only accepts time series with more than 20 data. 

\begin{figure}
    \centering
	\includegraphics[width=0.45\columnwidth]{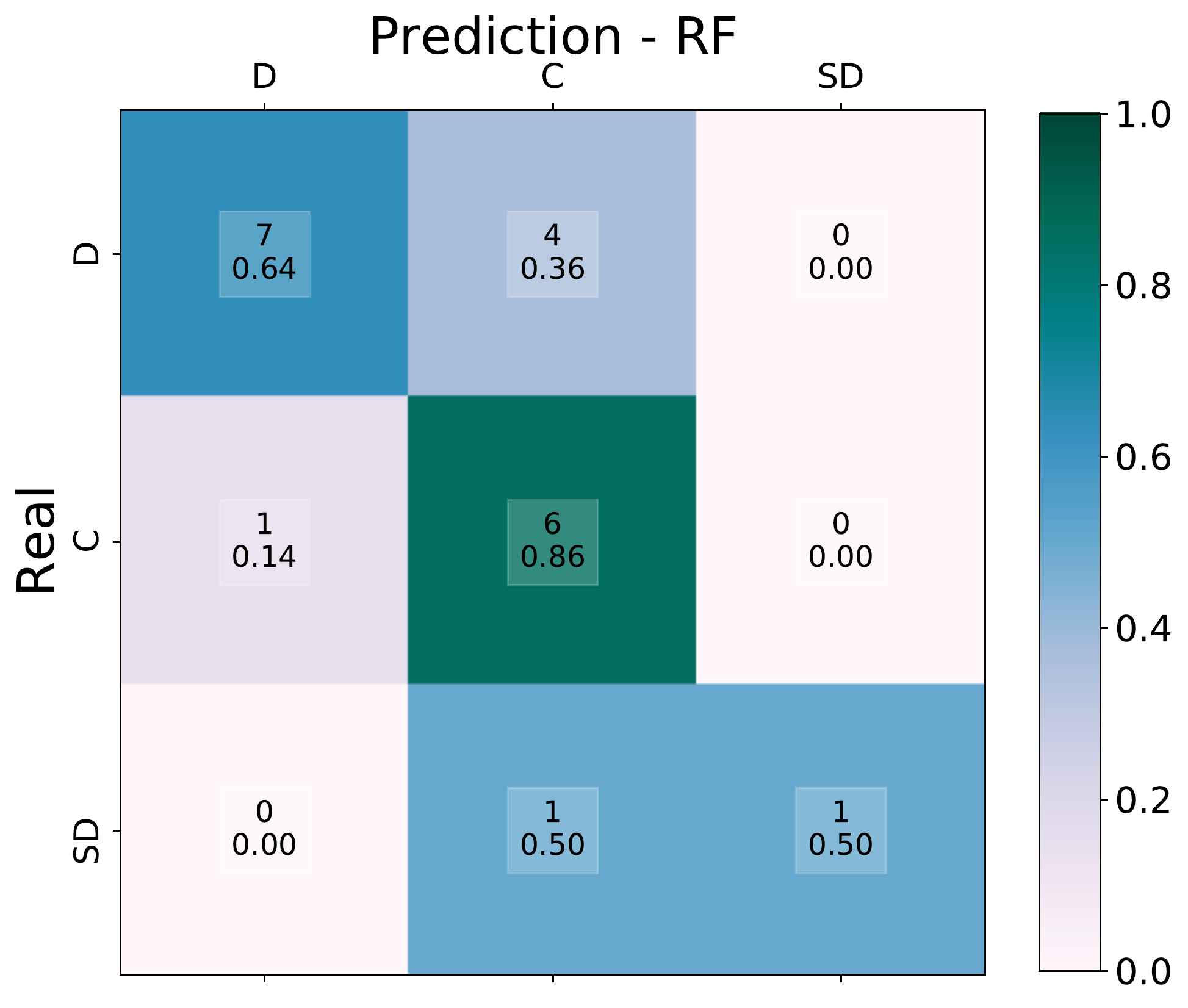}
    \includegraphics[width=0.45\columnwidth]{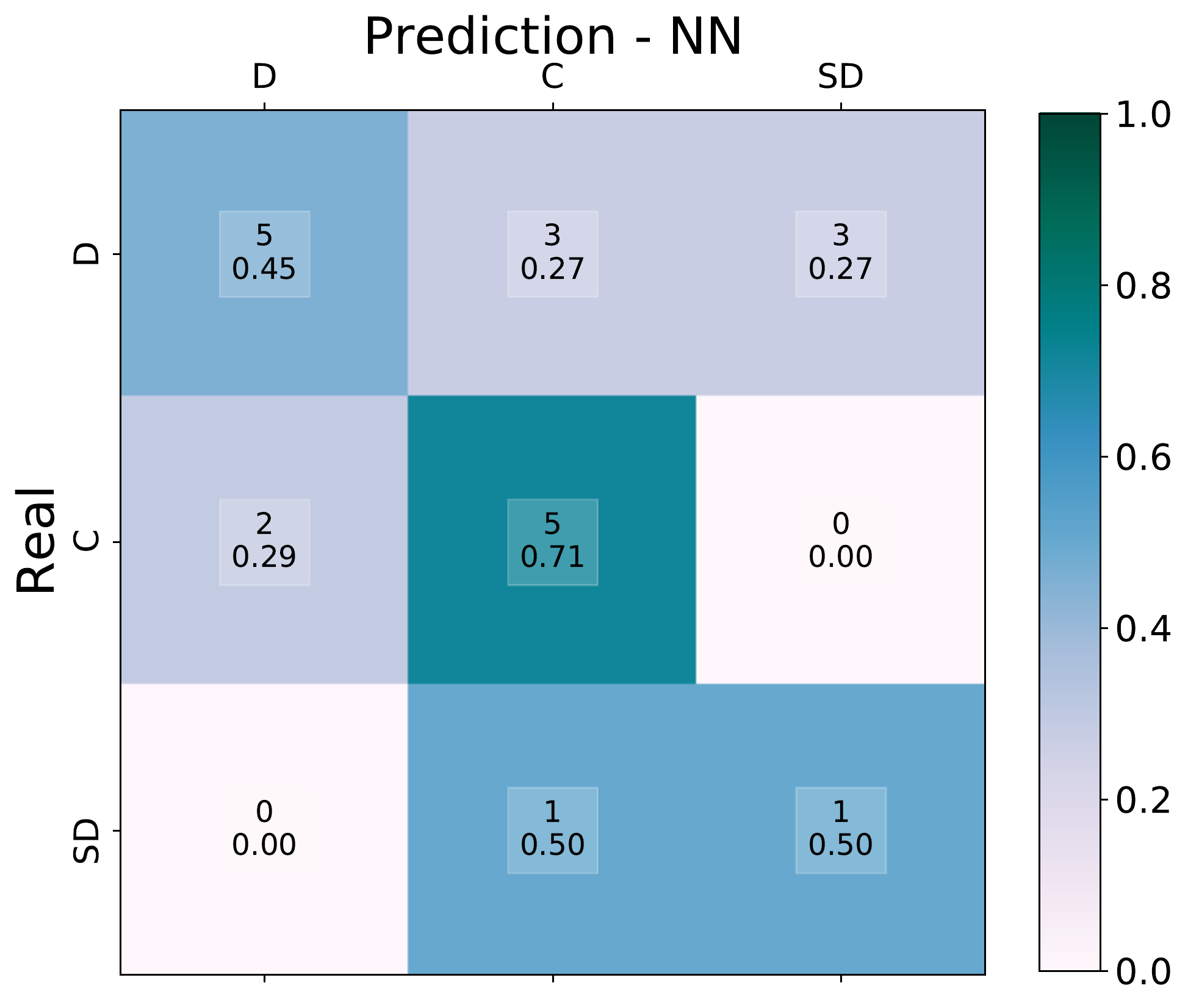}
	\includegraphics[width=0.45\columnwidth]{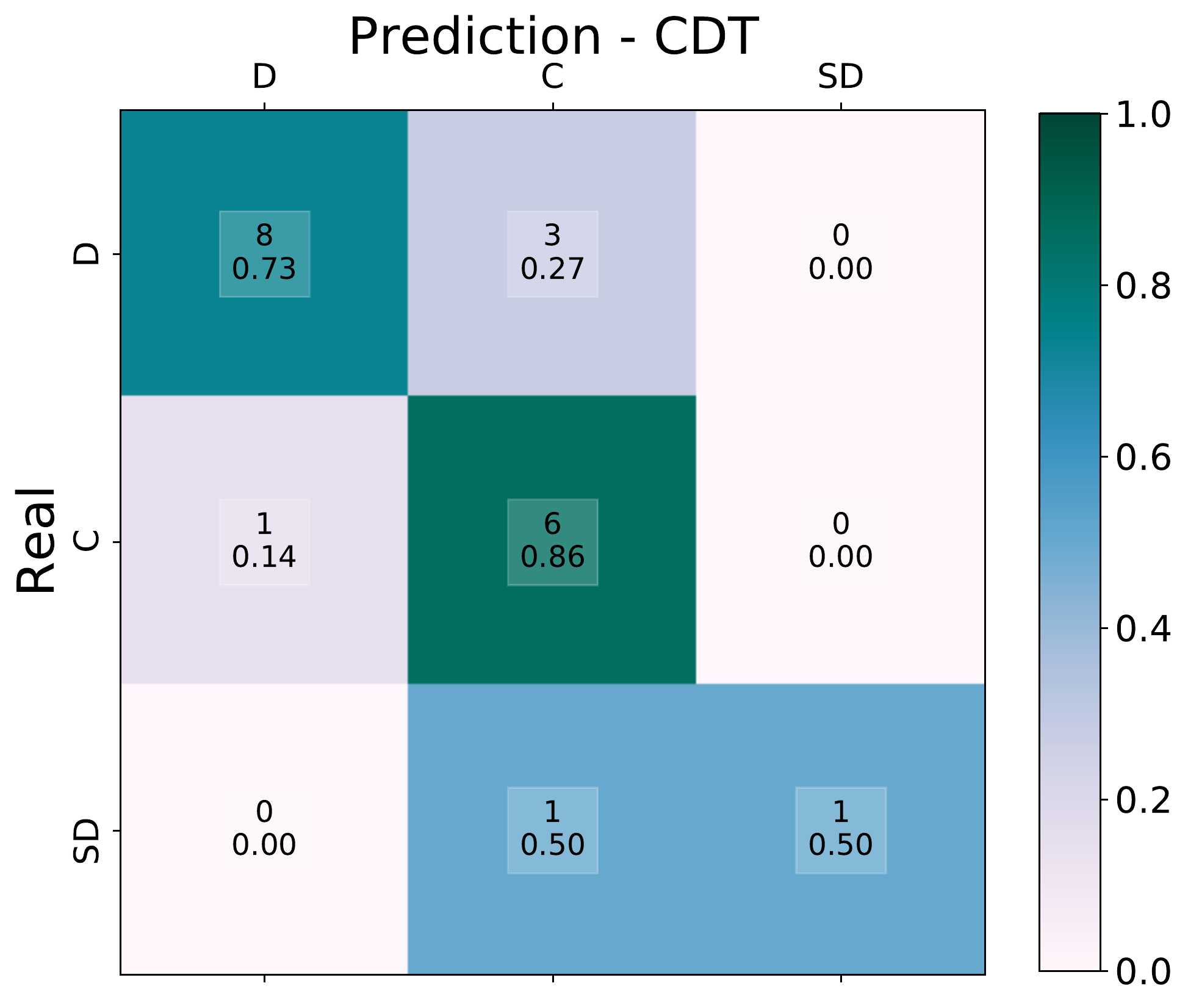}
    \caption{Confusion matrix visualisation of the performance of the test set with a training   set of 134 EBs.  The cells indicate the fraction of elements in rows that are classified with the labels in the columns, so that the elements in the diagonal are correctly classified. {\it Left upper panel}: performance of the RF model. {\it Right upper panel}: performance of the NN model. {\it Lower panel}: performance of the CDT model.}
	\label{fig:NN_RF_CDT}
\end{figure}

We then classified 58 EBs from the set of 38538 variable objects and compared  these classifications with the automatic classifications, the number of hits for RF was 14, NN 1 hit and the CTD model got 32 hits. Subsequently, we included the 58 EBs  to the set of 96, leaving a total of 154 EBs  whose class balance is shown in  Table~\ref{tab:d3_d2}. The 58 EBs  were included in the training  set and the models were run again, the evaluation of which on the test set results in an improvement of the classification of the SD EBs  (see Figure~\ref{fig:NN_RF_CDT}). 

\begin{figure}
	\includegraphics[width=0.45\textwidth]{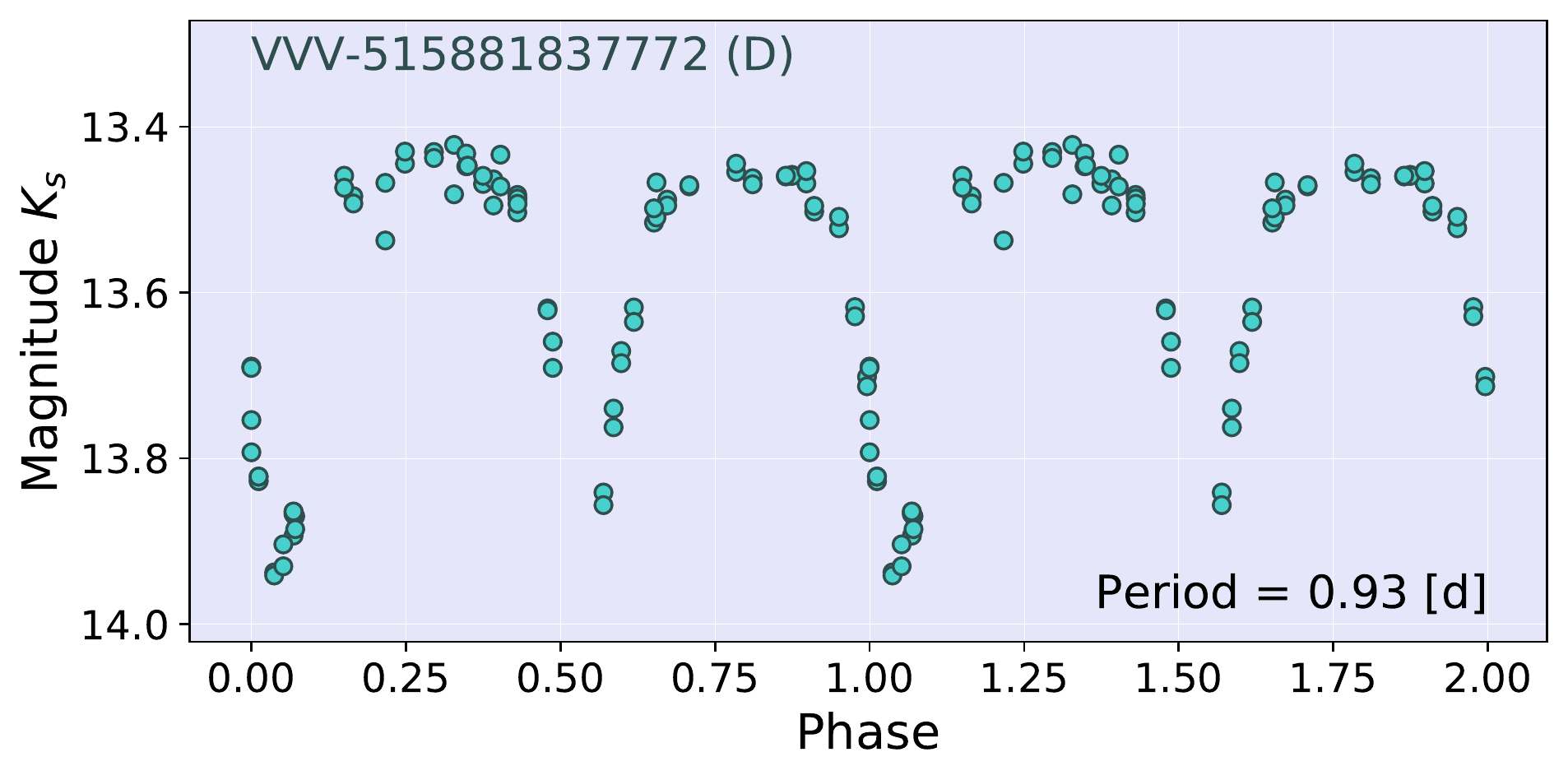}
	\includegraphics[width=0.45\textwidth]{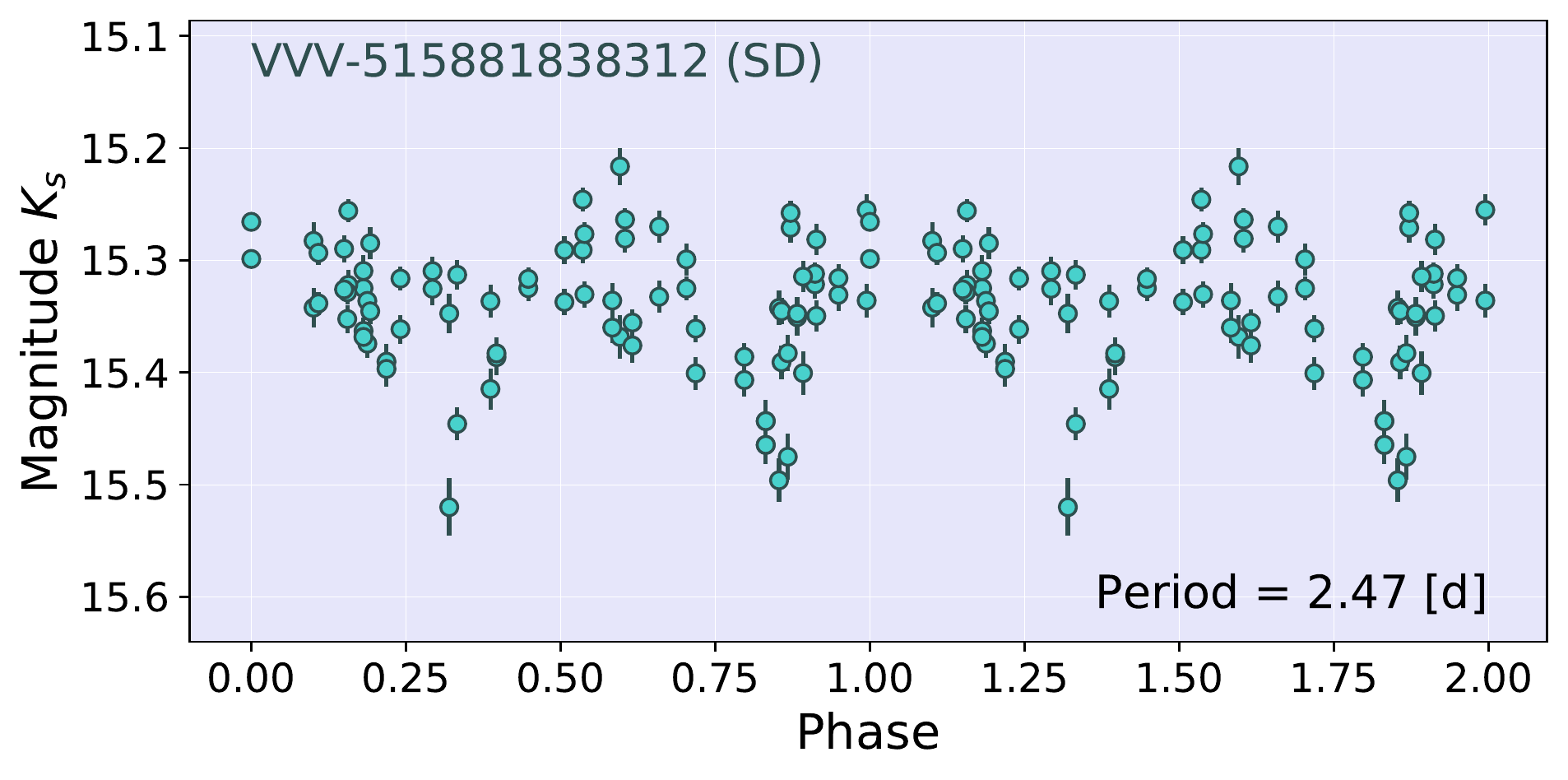}
	\includegraphics[width=0.45\textwidth]{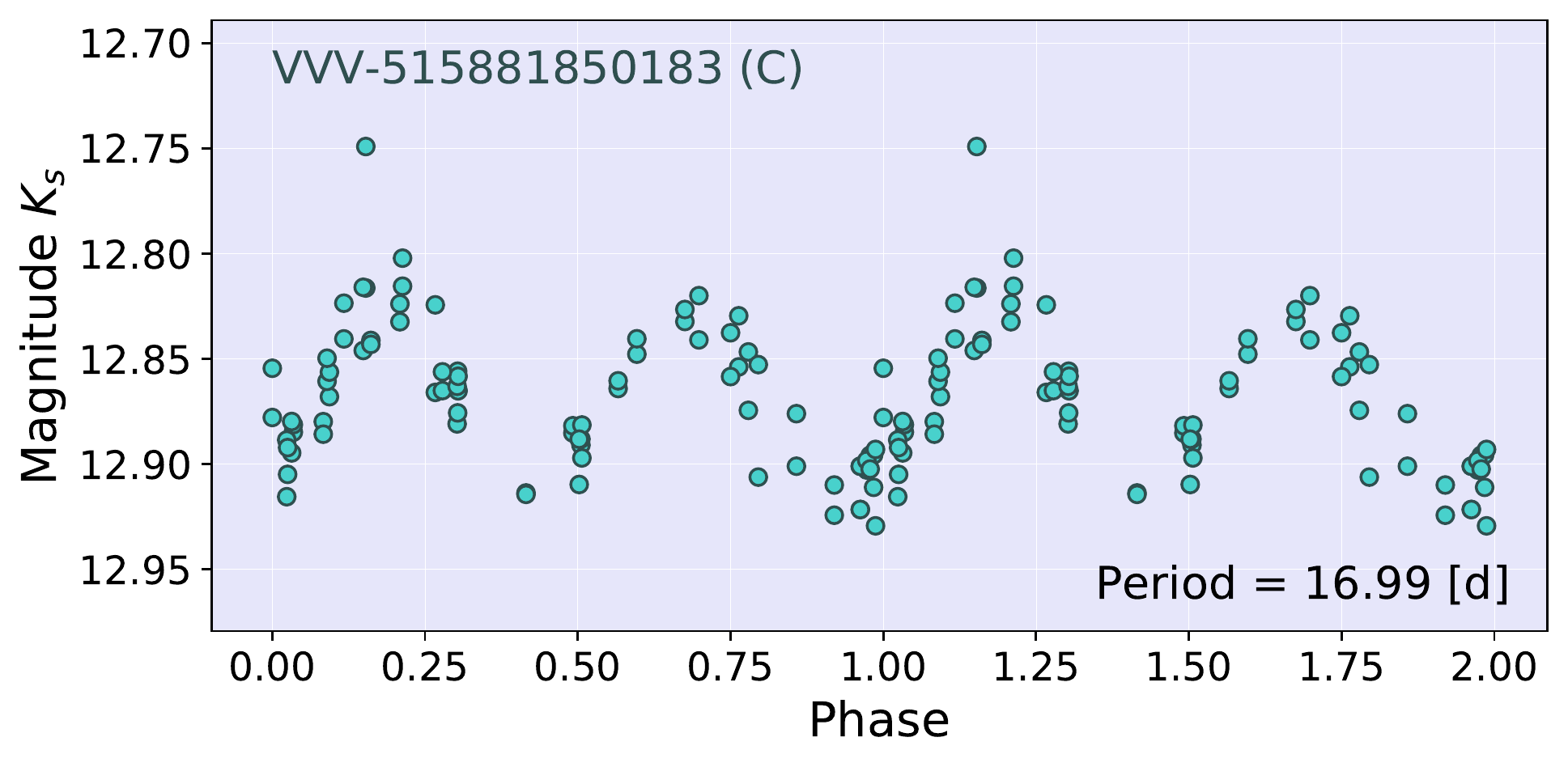}
    \caption{The light curves of the three EBs  wherein the primary minimum is located at zero. In the top panel, we show the light curve of a D-type  EB . In the middle and bottom panels, we show the light curves of SD and C-type EBs , respectively.}
	\label{fig:fit_D_SD_C_CL_d078}
\end{figure}

Then, 64 EBs from the set of variable objects were again classified by visual inspection and compared with the classification of the new models  . In this case, the number of hits for the RF and NN models was quite low, while CDT had 54 hits, i.e. approximately 80 per cent of the 64 EBs classified by visual inspection. The class balance of the 64 EBs plus the 134 EBs  is shown in Table~\ref{tab:d3_d2}, while the three light curves of D, SD and C type EBs are displayed in Figure~\ref{fig:fit_D_SD_C_CL_d078}.

According to these results, we concluded that the most reliable model is the CDT trained with the 96 EBs from the \citet{Gramajo2020} catalogue plus the 198 EBs obtained in this work, which have 35 features of the light curves selected by the MI  method and generated with \textsc{feets}.

\newcommand{\mybar}[4][1.5cm]{
\begin{tikzpicture}
\node[minimum width=#1] (top) {};
\node[fill=white, minimum width={(#2/#3)*#1},
      below right=-10pt, align=left]{#4};
\end{tikzpicture}
}
\newcommand{\mybarr}[4][1.5cm]{
\begin{tikzpicture}
\node[minimum width=#1] (top) {};
\node[fill=white, minimum width={(#2/#3)*#1}, below right=-8pt, align=right]{#4};
\end{tikzpicture}
}

\begin{table}
\centering
\begin{tabular}{lccc}
\hline
\hline
Type & Initial D & Clasif 1 & Clasif 2\\
\hline
\hline
D & 49 & 56 & 21\\

C & 35 & 50 & 19\\

SD & 12 & 28 & 24\\
\hline
\end{tabular}
\caption{Balance of the EBs   classes. The second column presents the balance of the 96 EBs,  the third the balance of 58 EBs   classes and the fourth shows the balance of the last visually classified dataset with 64 EBs. }
\label{tab:d3_d2}
\end{table}

\section{Conclusions}\label{sec:conclusions}

We developed a supervised ML model for the classification of EBs  into detached, semi-detached and contact classes in the VVV data. The construction of the model was made based on a catalogue of 100 EBs  determined in the VVV and classified object-by-object through the light curves using the \textsc{phoebe} software, plus periods determined by applying different methods, the generation of light curve features of each class through \textsc{feets} and the calculation of the magnitude difference between the primary and secondary maxima of the light curves.

By implementing statistical methods, we found that some components of the Fourier decomposition and parameters for estimating time scales are much less important in determining the classes of EBs.  Moreover, the combination of different sets of features selected according to their high entropy in relation to the classification reveals that only 35 features of shape, statistics and the LS period with which the systems were determined are sufficient to obtain a good performance in the classification of the EBs   of the VVV study. This is due to the fact that they have well differentiated distributions for the D and C classes. We also found that the distributions of the SDs are similar to those of the C systems for the most part but some features of the SD light curves are similar to those characterising the light curves of the D-type EBs. 

This set of features performs well for the usual supervised ML models with an imbalanced dataset, i.e., a higher number of examples of D-type EBs   compared to the number of examples of C-type EBs   and the minority class SD. An equal performance can be obtained with balanced datasets with increased minority classes, C and SD, as well as datasets with increased minority classes and a small decrease in class D. 

Usually the classification performance of variable stars, particularly EBs,  in a VVV tile has good performance with supervised learning models. In our study, that performance can be achieved with approximately one hundred EBs,  a small number of examples and the selection of the number of examples and features that enter the ML models. However, on examining the results, we found that this good performance is due to the good classification in the D and C classes but not so in the SD class. 

In order to achieve a good performance in the classification of the three EBs   classes, we developed a Compound Decision Tree (CDT) model, based on an algorithm that combines four supervised learning models with a voting system for EBs   classification. CDT starts by classifying classes C and D. If the classification is D, then the input will be classified into D and SD and if the classification is C, the input will be classified into C and SD. This model improved the classification performance of the D and C EBs   in the test set and therefore lowers the probability of misclassification for SD-type EBs   in this set.

When we used CDT for the classification of a new sample of EBs   in the d078 tile without examples of the systems in this tile in the model training   and compared the results with visually made classifications of 58 EBs,  CDT scores 32 hits outperforming models such as RF or NN. Subsequently, including the 58 EBs   in the training   set leads to an improvement in the classification of the SD-type EBs,  compared to a new one of 64 visually inspected systems. The CDT model obtains approximately 80 per cent of the visually performed classification, outperforming by far each of the ML models that comprise it.

Therefore, to classify EBs,   our CDT model needs few features of the light curves.
Thus, the classification of EBs   in a new tile with CDT outperforms ML models such as NNs or RFs.
The classification performance of EBs   is not only good for C- and D-type EBs   but also for SD-type EBs,  which are generally not well classified by supervised ML models dedicated to variable source classification.
Moreover, the CDT model is able to generate EB  catalogues in another VVV mosaic with a reliability of about 80 per cent. Its performance improves rapidly as soon as a few EB  examples are added to the mosaic to be classified.


\section*{Acknowledgements}


This work was partially supported by the Consejo Nacional
de Investigaciones Científicas y Técnicas (CONICET, Argentina) and the Secretaría de Ciencia y Técnica, Universidad Nacional de Córdoba, Argentina.
+ VVV DATA.
This research has made use of NASA’s Astrophysics Data System.

We gratefully acknowledge the use of data from the ESO Public Survey program IDs 179.B-2002 and 198.B-2004 taken with the VISTA telescope and data products from the Cambridge Astronomical Survey Unit. D.M. gratefully acknowledges support by the ANID BASAL projects ACE210002 and FB210003, and Fondecyt Project No. 1220724.

\section*{Data Availability}
 The data underlying in this article are available on request to the
corresponding author.

\bibliographystyle{mnras}
\bibliography{Bibliography} 

\bsp	
\label{lastpage}
\end{document}